\documentclass[12pt]{article}

\usepackage{psfig,epsfig}
\usepackage{psfrag}
\usepackage{amssymb}
\usepackage{amsmath}

\newcommand{\ba}{\begin{eqnarray}}
\newcommand{\ea}{\end{eqnarray}}
\newcommand{\nn}{\nonumber}

\newcommand{\delbar}{\overline{\partial}}
\newcommand{\Tr}{{\rm Tr\,}}
\newcommand{\tr}{{\rm tr\,}}

\newcommand{\calC}{{\cal C}}

\newcommand{\IZ}{\mathbb{Z}}
\newcommand{\IC}{\mathbb{C}}
\newcommand{\IP}{\mathbb{P}}

\newcommand{\IR}{\mathbb{R}}

\newcommand{\re}{{\rm Re \,}}

\newdimen\tableauside\tableauside=1.0ex
\newdimen\tableaurule\tableaurule=0.4pt
\newdimen\tableaustep
\def\phantomhrule#1{\hbox{\vbox to0pt{\hrule height\tableaurule width#1\vss}}}
\def\phantomvrule#1{\vbox{\hbox to0pt{\vrule width\tableaurule height#1\hss}}}
\def\sqr{\vbox{%
  \phantomhrule\tableaustep
  \hbox{\phantomvrule\tableaustep\kern\tableaustep\phantomvrule\tableaustep}%
  \hbox{\vbox{\phantomhrule\tableauside}\kern-\tableaurule}}}
\def\squares#1{\hbox{\count0=#1\noindent\loop\sqr
  \advance\count0 by-1 \ifnum\count0>0\repeat}}
\def\tableau#1{\vcenter{\offinterlineskip
  \tableaustep=\tableauside\advance\tableaustep by-\tableaurule
  \kern\normallineskip\hbox
    {\kern\normallineskip\vbox
      {\gettableau#1 0 }%
     \kern\normallineskip\kern\tableaurule}%
  \kern\normallineskip\kern\tableaurule}}
\def\gettableau#1 {\ifnum#1=0\let\next=\null\else
  \squares{#1}\let\next=\gettableau\fi\next}

\begin{document}
\noindent
\begin{titlepage}

\begin{center}
\hfill IHES/P/08/60

%\hfill hep-th/yymmnnn\\ 
\vskip 2cm {\Huge Phase space polarization and the topological\\ \vskip 0.2cm string: a case study}
\vskip 1cm {Amir-Kian Kashani-Poor\footnote{kashani AT ihes.fr}}\\ \vskip 0.5cm
{\it Institut des Hautes \'Etudes Scientifiques\\
Le Bois-Marie, 35, route de Chartres \\
91440 Bures-sur-Yvette \\
France}
\end{center} 
\vskip 1.5cm

\begin{abstract}
We review and elaborate on our discussion in hep-th/0606112 on the interplay between the target space and the worldsheet description of the open topological string partition function, for the example of the conifold. We discuss the appropriate phase space and canonical form for the system. We find a map between choices of polarization and the worldsheet description, based on which we study the behavior of the partition function under canonical transformations.

\end{abstract}

\end{titlepage}
\newpage

\section{Introduction}
Just as in physical string theory, topological string amplitudes were originally  defined genus by genus from a worldsheet formulation. Much progress has been made in the meantime in understanding these theories from a target space point of view, which often provides better computability than the worldsheet approach. Unlike the worldsheet description, the target space approach should in principle provide a non-perturbative definition of the theory. The object which is conventionally referred to as the topological string partition function is indeed a partition function for this target space theory, i.e. the path integral of the action over target space fields. On a target space with boundary, the path integral yields a functional on the boundary values of the fields. In the Hamiltonian formalism, we must choose a polarization on field space, and only specify half of the fields on the boundary. The functional thus defined is a wave function, and transforms as a wave function must under a change of polarization. Though no target space path integral enters into the definition of the worldsheet partition function, it must exhibit this wave function behavior if a dual target space description indeed exists.

Following the treatment in \cite{me}, we study below how the choice of polarization manifests itself in the open topological string partition function, in the example of the resolved conifold.  We find that the proper choice of canonical variables depends on such data as brane placement, the distinction between branes and antibranes, and the choice of K\"ahler cone. Having mapped different partition functions to choices of polarization, we have a prediction for the behavior of the partition function under canonical transformations. Performing the integral transforms explicitly matches these predictions, but only up to coefficients and argument shifts. We interpret these mismatches as indicative of non-perturbative effects that are missed by the worldsheet treatment.

\section{The A-model perspective} \label{A}
\subsection{The geometry}
\begin{figure}[h]
\psfrag{x}{$|X_1|^2$}
\psfrag{y}{$|X_2|^2$}
\psfrag{z}{$|X_3|^2$}
\psfrag{w}{$|X_4|^2$}
\begin{center}
\epsfig{file=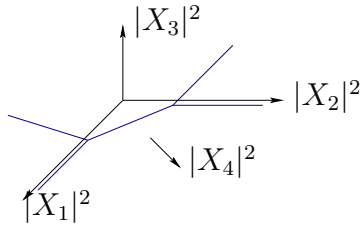, height=1in}
\end{center}
\caption{\small Conifold as $T^3$ fibration over manifold with corners. \label{fig:conifoldasfibr}}
\end{figure}

We consider the A-model on the resolved conifold ${\cal O}(-1) \oplus {\cal O}(-1) \rightarrow \IP^1$. It can be obtained via a symplectic quotient construction with moment map
\ba \label{momentcfld}
|X_1|^2 + |X_2|^2 - |X_3|^2 - |X_4|^2 &=& t \,.
\ea
The zero set of this equation in $\IC^4$ is depicted in Fig. \ref{fig:conifoldasfibr}. The geometry is toric, with the obvious $(\IC^*)^3$ action. Regarded as a toric fibration, one, two, three 1-cycles respectively of the fiber degenerate on each face, edge, corner of the base. In the following, we will sometimes depict this geometry simply by indicating the edges and corners (or vertices) along which the fiber degenerates. We will refer to this locus as the toric skeleton.

\subsection{A-branes} \label{s:Abranes}
A-branes wrap Lagrangian submanifolds and carry a flat connection. A distinguished class of Lagrangian submanifolds in toric varieties can be obtained as $T^2$ fibrations over lines in the geometry of Fig. \ref{fig:conifoldasfibr} (linear in the coordinates $|X_i|^2$) ending on the toric skeleton \cite{AV}. The slope of the lines must be correlated with the embedding of the $T^2$ fiber of the Lagrangian into the $T^3$ fiber of the ambient geometry in order to satisfy the Lagrangian condition (with regard to the symplectic form inherited from $\IC^4$, $\omega = \sum_i d|X_i|^2 \wedge d\theta^i$). These Lagrangians have the topology $\IR^2 \times S^1$: of the two $S^1$ fibers at infinity, one degenerates on the toric skeleton.

The moduli space of branes wrapping these distinguished special Lagrangian submanifolds is locally a cylinder: it is given by the position along an edge, encoded in the area $e^{-r}$ of a holomorphic disc ending on the brane, together with a phase given by the Wilson loop of the brane gauge field around the $S^1$, specifying the flat bundle over the Lagrangian,
\ba
x = e^{-r}\, \tr P \exp \oint A \,. \label{complexified}
\ea
$\tr$ here specifies the trace in the fundamental representation - for the Abelian bundles that we will be considering, the trace and path ordering are of course trivial.

At a vertex, the entire fiber shrinks to zero, and we hence cannot naively follow the brane through the vertex from one edge to another. From the classical A-model point of view, the complete moduli space hence appears disconnected, consisting of a copy of a cylinder per edge \cite{hori}. However, as common in the A-model, we should expect corrections to this description quantum mechanically. Indeed, as we review below, a B-model analysis shows that in the quantum corrected moduli space, the cylinders are in fact joined together smoothly to form a 4-punctured sphere \cite{KatzKlemm}.

To completely specify the open A-model on non-compact geometries, an additional integer must be specified for each non-compact brane in the geometry. This integer was referred to as a framing choice in \cite{AKV}, as this is what it corresponds to in the target space description of the open topological string given by Chern-Simons theory.

\subsection{The target space description} \label{ATS}
In \cite{cswitten}, Witten provides a target space description of the open topological string in the A-model as a Chern-Simons theory living on the Lagrangian submanifold wrapped by the brane, modified by instanton corrections,. Given a Lagrangian with boundary, the partition function of this target space theory should behave as a wave function on the phase space given by the field configurations restricted to the boundary. We interpret the non-compact Lagrangians introduced in the previous subsection as solid tori with a boundary $T^2$ at infinity. The phase space of conventional Chern-Simons theory on a solid torus is given by the holonomy of the gauge field around its two cycles \cite{Elitzur}. In a Hamiltonian framework, only one of these two variables is fixed as a boundary condition when evaluating the path integral of the target space action. By standard arguments in quantum mechanics, different choices of cycle yield partition functions which are related by canonical transformations, as has been explicitly worked out in \cite{Elitzur}.

Witten's target space proposal for the open topological A-model differs from conventional Chern-Simons theory in two regards: the presence of instanton corrections, and the complexification of the fields, as in Eq. (\ref{complexified}), to encode moduli of the Lagrangian submanifolds in addition to the connection on the brane. While the instanton corrections take place away from the boundary and hence do not enter in determining the phase space of the theory, the complexification of the fields leads to a complexification of phase space which will play a central role in the analysis below.

\section{The B-model perspective} \label{s:tmg}
\subsection{The geometry}
In the Hori-Vafa prescription, the superpotential of the mirror Landau-Ginzburg theory to the A-model on the conifold is given by
\ba
W &=& \sum_{i=1}^n e^{-Y_i} \,, \nn
\ea
with $\re Y_i = |X_i|^2$, and the linear constraint
\ba
Y_1 + Y_2 - Y_3 -Y_4 &=& t  \label{toricdata}
\ea
among the $Y_i$ reflecting the toric data \cite{HV}. As far as holomorphic data is concerned, we can instead consider the B-model on the geometry 
\ba
\sum_{i=1}^n e^{-Y_i} &=&UV \,,\nn
\ea
with $e^{-Y_i}$ homogeneous coordinates on $\IC\IP^{3}$ and $U$ and $V$ sections of an appropriate line bundle over this projective space \cite{HIV}. Upon solving Eq. (\ref{toricdata}) and restricting to an affine patch, the geometry is governed by an equation of the form
\ba
H(w,y) &=& uv  \,,
\ea
with $H(w,y)$ a polynomial in $e^{\pm w}, e^{\pm y}$.

\subsection{B-branes}
B-branes, at the level of precision required here, wrap holomorphic submanifolds and carry holomorphic vector bundles.

The mirrors to the Lagrangian branes ending on the toric skeleton discussed in section \ref{A} wrap holomorphic curves given by \cite{AV}
\ba
uv &=& 0 \,,\nn\\
H(w_0, y_0) &=&  0  \,.
\ea
The moduli space of these branes is coordinatized locally by the coordinates $w_0, y_0$, which are constrained to lie on the Riemann surface determined by $H$. In terms of homogeneous coordinates on $\IC \IP^3$, this Riemann surface is given by the equation
\ba \label{eqrs}
e^{-Y_1}+e^{-Y_2}+e^{-Y_3}+e^{-Y_4}=0 \,,
\ea 
and Eq. (\ref{toricdata}), homogeneous in the coordinates $y_i = e^{-Y_i}$. These equations describe a sphere. As the physical variables describing the B-model are the cylinder variable $Y_i$, the correct moduli space requires puncturing the sphere in the following four points: 
\ba
P_1 : (1:0:-1:0)\,, P_2:(1:0:0:-1)\,, P_3:(0:1:0:-1)\,, P_4:(0:1:-1:0)\,. \nn
\ea
At each of these, the real part of two of the coordinates $Y_i$ is going off to infinity. In the $Y_i$ coordinates, this 4-punctured sphere naturally appears as 4 cylinders which are joined smoothly. We easily identify these cylinders as the external edges of the toric diagram describing the resolved conifold. As promised, the moduli space of branes that arises in the B-model setup hence appears to connect the disconnected moduli space that arose in the A-model picture.
  	
\subsection{The target space theory} \label{s:BTST}
The action of holomorphic Chern-Simons theory, the proposed target space theory of the open topological B-model, is given by
\ba
\int \Omega \wedge \Tr ( A \wedge \delbar A + \frac{2}{3} A\wedge A \wedge A)  \,,
\ea
where $A$ is a holomorphic connection \cite{cswitten}. The appropriate description of non-compact B-branes is conjectured to be given by the dimensional reduction of this action onto the worldvolume of the brane \cite{AV}. Restricting to an affine patch and an infinitesimal neighborhood of the curve $\calC$ yields the reduced action
\ba
\int_\calC \frac{du d\bar{u}}{u} w \delbar_u y \,,
\ea
with $u$ the coordinate along the brane worldvolume $\calC$. The punchline of this analysis is that $w$ and $y$ are conjugate coordinates from the point of view of the brane target space action. The appropriate symplectic form on this patch of phase space is therefore
\ba
\omega = \pm dw \wedge dy \,. \label{symplectic}
\ea
Note that we are proposing a 2 complex dimensional phase space with a $(2,0)$ form as holomorphic symplectic form, and our wave functions are to depend holomorphically on one complex variable. This is a reflection of the complexification of the open string modulus on the A-model side pointed out in subsection \ref{ATS}. 

The equation $H(w,y)=0$ has not entered into our analysis up to this point. We will interpret this equation, following \cite{ADKMV}, as a constraint to be imposed on the topological string partition function {\it after} quantization. More precisely, upon a choice of canonical variables $(p(w,y),x(w,y))$, we obtain a differential operator from $H(w,y)$  (up to normal ordering ambiguities to which we return below) by rewriting it in terms of $p,x$, mapping $p \mapsto g_s \partial_x$ and interpreting $x$ as a multiplication operator. We then require this operator to annihilate the partition function. \cite{ADKMV} motivates this procedure by considering the Landau-Ginzburg description of the B-model with superpotential $W=H(w,y)$, and imagining dynamically turning off $W$. Imposing Eq. (\ref{toricdata}), but not Eq. (\ref{eqrs}) on $\IC \IP^3$ yields a 2 complex dimensional space $\tilde{M}$, which we interpret, upon puncturing, as the holomorphic phase space $M$ of our system.

In the homogeneous coordinates of $\IC \IP^3$, $\tilde{M}$ is given by
\ba
y_1 y_2 &=& y_3 y_4 e^{-t} \,.
\ea
Consider first the point $t=0$ in complex structure moduli space. By setting
\ba
y_1 = a b \;,\;\; y_2 = c d \;,\;\; y_3 = ad \;,\;\; y_4=bc \,, \label{newvar}
\ea
we can identify this space with $\IC \IP^1 \times \IC \IP^1$, coordinatized as $(a:c) \times (b:d)$. We obtain $M$ by removing the points where any of the coordinates $a,b,c,d$ vanish, obtaining $\IC^* \times \IC^*$. While one may at this point be tempted to forget about $\tilde{M}$ and introduce affine coordinates on $M$, we will see that the description in terms of $\{a,b,c,d\}$, corresponding each to one of the four punctures of the sphere, is best suited for assigning polarizations to punctures. The proposal for a holomorphic symplectic form on $M$ consistent with Eq. (\ref{symplectic}) is now
\ba
\omega &=& \pm \;d \log \frac{a}{c} \wedge d \log \frac{b}{d} = \pm d \xi \wedge d \eta \,, \label{symplectic form conifold}
\ea
where
\ba
\xi &=& \log \frac{y_1}{y_4} = \log \frac{y_3}{y_2} \,, \nn \\
\eta &=& \log \frac{y_1}{y_3} = \log \frac{y_4}{y_2} \,.
\ea
For arbitrary $t$, this discussion goes through after a rescaling of the identification of $y_i$ and $a,b,c,d$, eg. by retaining Eqs. (\ref{newvar}) up to
\ba
y_2 &=& e^{-t} cd \,.
\ea
This does not effect the choice of symplectic form, but introduces a shift of the canonical variables by $\pm t$ in certain instances, as we will develop below.

\section{The partition function as wave function} \label{s:wave}
\subsection{Choice of polarization and the worldsheet} \label{s:bpcp}
\begin{figure}[h]
\centerline{\epsfig{file=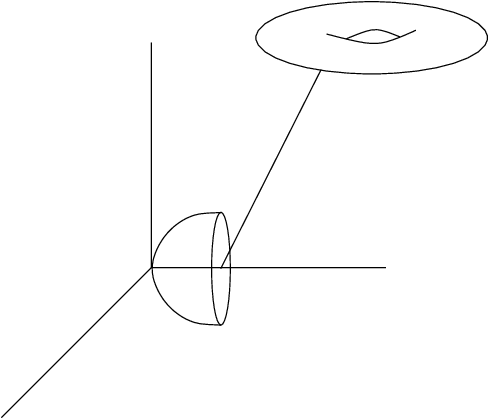, height=1in}}
\vspace*{8pt}
\caption{\small The boundary torus at infinity vs. the degenerate torus on which the worldsheet ends. \label{fig:t vs w}}
\end{figure}

We have now laid the necessary ground work to tackle the question we raised in the introduction: identifying the dependence of the partition function on the choice of polarization of phase space, which arises naturally in the target space description of the theory, from the worldsheet point of view. We begin by considering the A-model. It is natural to identify the dependence of the partition function on the modulus defined in Eq. (\ref{complexified}) with the boundary condition dependence in the target space description, reviewed in section \ref{ATS}. Indeed, both descriptions require the choice of a cycle of the torus fiber of the Lagrangian. However, the relevant torus for the target space description sits at infinity, whereas the relevant torus for the worldsheet is situated at the center of the solid torus and is hence degenerate. The two tori are sketched in figure \ref{fig:t vs w}. 

The only natural choice for defining the worldsheet modulus is a non-contractible cycle of the solid torus. It hence would appear that some polarizations are not accessible from the worldsheet description of the theory. However, based on the  B-model description of the setup, we developed an understanding of the moduli space of branes as a 4-punctured sphere, i.e. four {\it joined} cylinders, in the previous section \cite{AV}. We can therefore determine the open topological string partition function for a brane close to a given puncture, corresponding to a Lagrangian ending on the corresponding edge, and consider following the partition function as we move along moduli space to a different puncture, i.e. to a different edge of the toric diagram from the A-model point of view. This procedure therefore allows us to obtain the open topological string partition function expressed in the unnatural coordinate corresponding to the complexified Wilson loop around a contractible cycle (since the functional form of the partition function does not change as we move between punctures, one can somewhat glibly say that different brane placements in the open topological string partition function are related via changes of polarization of phase space) \cite{ANV}.

Complementing brane placement by choice of framing and K\"ahler cone, we will find in the remainder of this section that we can uniquely map the choice of polarization to the worldsheet theory.\footnote{To be precise, while the framing must be specified in determining the open partition function on non-compact manifolds and hence a worldsheet manifestation must exist, it has not been identified to date (but see \cite{katz}).} 

\subsection{The partition functions} \label{s:pfct}
\begin{figure}[h]
\psfrag{a}{1}
\psfrag{b}{2}
\psfrag{c}{3}
\psfrag{d}{4}
\begin{center}
\epsfig{file=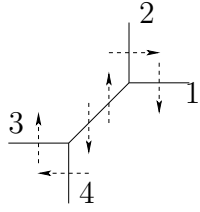, height=1in}
\end{center}
\caption{\small Conifold with fiducial framing. \label{fig:conifold}}
\end{figure}

The open string partition function in the presence of a single brane wrapping a Lagrangian submanifold of the form discussed in subsection \ref{s:Abranes} takes the form
\ba
Z(x) &=& \sum_R Z_R e^{-|R|x} \,,
\ea 
with $Z_R$ the partition function as calculated in the topological vertex formalism of \cite{AKMV}. A representation $R$ is assigned to the appropriate edge of the toric diagram. The sum ranges over all Young tableaux consisting of a single row (including the empty tableau), $|R|$ corresponds to the number of boxes in the tableaux, and the open string modulus $x$ is as defined in Eq. (\ref{complexified}). The partition function for an antibrane (in the sense of \cite{vafanti}) is
\ba
Z(x) &=& \sum_R (-1)^{|R|} Z_{R^t} e^{-|R|x} \,.
\ea 
For the conifold, we label the external edges of the toric diagram from $1$ to $4$ as in Fig. \ref{fig:conifold}. The partition functions are then given by \cite{me}
\ba
Z^{1,3(k)}(x) &=& \sum_{r=0}^{\infty}  \frac{\prod_{i=1}^{r} (1-Qq^{1-i})}{[r]!} q^{\frac{1}{4} (r-1)r(1-2k)} (-1)^{rk} e^{-rx} \,, \nn \\
Z^{2,4(k)}(x) &=&\sum_{r=0}^{\infty} (-1)^r \frac{\prod_{i=1}^{r} (1-Qq^{i-1})}{[r]!} q^{-\frac{1}{4} (r-1)r(1+2k)}  (-1)^{rk} e^{-rx} \,.
\ea
The superscript on $Z$ indicates the edge on which the brane is ending, $Q= e^{-t}$ labels the exponential of the K\"ahler class, and $[r] = (q^{r/2} - q^{-r/2})$, $[r]!=\prod_{i=1}^r [r]$. The framing integer $k$ is defined relative to the framing depicted in Fig. \ref{fig:conifold}.

For the fiducial framing, the partition functions have a representation as infinite products
\ba
Z^{1,3}(x)&=& \prod_{i=0}^{\infty} \frac{1- q ^{i+\frac{1}{2}}e^{-x}}{1-Qq^{i+\frac{1}{2}}e^{-x}} \label{productrep13}\\
 &=& \exp[ \sum_{n=1}^{\infty} \frac{e^{-nx}}{n[n]}(1-Q^n)] \,, \nn
\ea
\ba
Z^{2,4}(x)&=& \frac{1}{Z^{1,3}(x)} \,.
\ea
The partition functions for anti-branes, which we denote as $Z^*$, are given by
\ba
Z^{*1,3(k)} &=& Z^{2,4(-k)} \,,\\
Z^{*2,4(k)} &=& Z^{1,3(-k)} \,.
\ea

\subsection{The assignment of canonical variables} \label{s:assignment}

Our goal is now to assign a choice of polarization of phase space to each external edge of the toric skeleton of the A-model geometry (hence each puncture of the B-model Riemann surface) that is natural from the point of view of the worldsheet. `Natural' here means that the canonical coordinate of this choice must map to a viable modulus from the worldsheet perspective, with its phase corresponding to a Wilson loop around a non-contractible $S^1$. As reviewed in subsection \ref{s:Abranes}, a choice of cycle of the solid torus which constitutes the worldvolume of the brane corresponds to a choice of (a product of) the four coordinates $X_i$ specifying the A-model geometry. Two of these do not vanish anywhere along the worldvolume of the brane, their phases hence yield viable $S^1$'s. These two coordinates are uniquely characterized by the property that they go to infinity along the edge on which the brane terminates, hence map to B-model variables $y_i$ that vanish at the corresponding puncture. This is the criterion which determines the choice of canonical coordinate \cite{ADKMV}.

Based on this discussion and the choice of symplectic form in Eq. (\ref{symplectic form conifold}), we arrive at the following procedure for assigning polarization in terms of the homogeneous coordinates $a,b,c,d$ on $\tilde{M}$ introduced in Eqs. (\ref{symplectic form conifold}). At each puncture, exactly one of these coordinates vanishes. The canonical coordinate is chosen appropriately; eg. at $P_1$, $c=0$, hence $x= - \xi$. The canonical momentum is then chosen in accordance with the symplectic form (\ref{symplectic form conifold}). At $P_1$, with the choice of sign $\omega = d \xi \wedge d \eta$, $p = \eta + n \xi$ (such that $\omega = dp \wedge dx$).

This simple picture must be refined due to the fact that the physical coordinates of the problem are the $Y_i$, not the $a,b,c,d$. As a consequence, the canonical coordinate at eg. $P_1$ is either $\log \frac{y_4}{y_1}$ or $\log \frac{y_2}{y_3}$. The choice is between the two variables that vanish at the puncture, in this case $y_2$ and $y_4$. In terms of the conifold geometry, this is roughly speaking a choice between (I) a base coordinate on $\IP^1$ or (II) a fiber (the two line bundles over $\IP^1$) coordinate. The distinction in terms of $\xi$ is a shift by $t$. For the canonical momentum, there is a natural choice: the quotient of the two variables $y_i$ that do not vanish at the puncture.

In table \ref{concor}, we list the polarization assignments to each puncture following the above discussion, for $n=0$ and $\omega = + d\xi \wedge d\eta$.

\begin{table}[h] 
\begin{center}
{\begin{tabular}{||c|c||c|c||c|c||c|c||}
 \multicolumn{2}{||c||}{$P_1$} & \multicolumn{2}{c||}{$P_2$} & \multicolumn{2}{c||}{$P_3$}&\multicolumn{2}{c||}{$P_4$} \\ \cline{1-8}
 $p_1$  &$x_1$ &$p_2$ & $x_2$ & $p_3$ & $x_3$ & $p_4$ & $x_4$ \\ \hline
 $\eta$ & $-\xi$ & $-\xi$ & $-\eta$ & $-(\eta-t)$& $\xi-t$ & $\xi-t$ & $\eta-t$ \\ \cline{1-8}
 $p$ & $x$ & $x$ & $-p$ & $t-p$ & $-t-x$ & $-t-x$ & $-t+p$ \\ 
%  II & $\log y_1/y_3$ & $-\log y_2/y_3$ & $-\log y_1/y_4$ & $-\log y_2/y_4$ & $ \log y_2/y_4$ & $-\log y_1/y_4$ & $ -\log y_2/y_3$ & $-\log y_1/y_3$ \\ \cline{2-9} 
%  & $p$ & $x$ & $t-x$ & $t+p$ & $-t-p $ & $t-x $ & $x$ & $-p$ 
\end{tabular}} 
\end{center}
\caption{Canonical variables for the conifold, with the canonical coordinate chosen along the fiber (II). \label{concor}}
\end{table}

We have arrived at a prescription for attributing a set of polarizations of phase space to each puncture of the B-model Riemann surface. As we will develop in the next subsection, it turns out that each of the remaining choices,
\begin{itemize}
\item the shift by $\pm t$ alluded to above,
\item the choice of sign of the symplectic form,
\item the choice of $n$ in $p \rightarrow p + nx$,
\end{itemize}
has an interpretation from the point of view of the worldsheet.

\subsection{The constraint equation}
We address the choices that remain in the puncture/polarization assignment with the help of the constraint equation (\ref{eqrs}): each choice of polarization maps the constraint to a different differential operator. Each partition function presented in subsection \ref{s:pfct} incorporates certain data (brane placement, brane vs. antibrane, etc.). By determining which partition function is annihilated by which operator, we can map this data to a choice of polarization.

Expressed in terms of the canonical variables of table \ref{concor}, the constraint equation takes the following forms,
\ba 
P_{1,3} &:& 1+ e^{-x}+e^{p}+e^{-x+p-t} = 0 \nn\\ 
P_{2,4} &:& 1+ e^{-x}+e^{-p}+e^{-x-p-t} = 0  \,.\label{rscon}
\ea
As the partition functions at $P_1, P_3$ and $P_2, P_4$ respectively coincide, it is a first test on the consistency of our setup that we obtain the same equations at the respective pairs. We find that the partition functions $Z^{1,2,3,4}$ are indeed annihilated by an appropriate constraint operator, once we take the following two points into account.
\begin{itemize}
\item Mapping an algebraic to a differential operator entails normal ordering ambiguities. In our case, due to $[p,x] = g_s$, they arise as powers of $q$. We adjust these as appropriate.
\item We have not been careful in keeping track of the imaginary parts of the variables $Y_i$. The correct constraint equations arise upon acting by $x,p \mapsto x+ i\pi,p + i \pi$ on Eqs. (\ref{rscon}).
\end{itemize}
We arrive at
\ba
(1- e^{p} - q^{-1/2} e^{-x} + q^{-1/2} e^{-x}Qe^{p}) Z^{1,3} &=& 0 \nn \,,\\
(1- e^{-p} - q^{1/2} e^{-x} + q^{1/2} e^{-x}Qe^{-p}) Z^{2,4} &=& 0 \,. \label{souped up constraints}
\ea

\paragraph{Phase space orientation:} Had we chosen the opposite sign of the symplectic form in equation (\ref{symplectic}), we would have identified the negative of the expressions listed in table \ref{concor} as canonical momenta. That $p \mapsto -p$ distinguishes between branes and antibranes is immediate upon inspection of the two equations (\ref{rscon}): they are exchanged by $p \mapsto -p$, consistent with $Z^{1,3} = Z^{2,4*}$. We hence observe that the choice between brane and antibrane is determined by the orientation of phase space.

\paragraph{Choice of K\"ahler cone:} The canonical coordinates corresponding to choice (I) and (II) introduced in the previous subsection are related by 
\ba
(p,x)_{II} &=& (p,x-t)_{I} \,.
\ea
Under this substitution, the two constraints (\ref{souped up constraints}) are essentially swapped, with $\tilde{Q}=e^t$ replacing $Q=e^{-t}$. We conclude that the choice of (I) vs. (II) in polarization corresponds to a choice of K\"ahler cone from the point of view of the worldsheet. By $Z^{1,3*}=Z^{2,4}$, the role of branes and antibranes are reversed with this choice.

\paragraph{Framing:} We can also understand the canonical transformation $T^k$, which adds multiples of $p$ to $x$, $x \mapsto x-kp$. Under this transformation, the constraint operator eg. for $Z^{1,3}$ becomes
\ba
1- e^{p} - (-1)^k q^{-1/2} e^{-x}e^{kp} + (-1)^k q^{-1/2} e^{-x}Qe^{(k+1)p} \,.
\ea
With the choice of normal ordering displayed, it annihilates the partition function $Z^{1,3(k)}$. The case $Z^{2,4}$ works analogously.
The correspondence between the integer ambiguity arising by adding multiples of $p$ to $x$ and the framing ambiguity was already developed in \cite{AV} in a related discussion.

\subsection{Transformation of the partition function} \label{s:transform}

We begin by studying how the operators $T$ and $S$ are represented on the Hilbert space of which the open string partition function is to be a wave function. Throughout this subsection, we are forced to be cavalier about pre-factors and shifts of $x$ by $t$ and $g_s$. We will have some comments in this regard in the following subsection, but a better understanding of the failure of the transformations studied here to yield exact results is still lacking.

Working out the kernel for the integral transform that implements a change of polarization on a wave function is a canonical exercise, yielding
\ba \label{intkernel}
K_A(x',x) &=& e^{\frac{1}{2g_s c} ( d x^2 - 2 xx' + a x'^2)}
\ea
for the $SL(2,\IZ)$ element
\ba
A =\left(
 \begin{matrix} 
      a & b \\
      c & d \\
   \end{matrix}
\right) \,.
\ea
The subtlety that arises in our setup is that the canonical variables are complex, the symplectic form is hence of type (2,0). This invalidates the canonical treatment for which the canonical variables are assumed to be self-adjoint. We pursue here a hybrid strategy: we use the integration kernels obtained from the canonical approach, but give ourselves leeway in the choice of integration contour \cite{ADKMV}.

Whenever the symplectic transformation has $d\neq0$, we interpret the integral as a Gaussian with the contour chosen to ensure convergence, and evaluate
\ba \label{mastergauss}
\int K_A(x',x) e^{-rx} &\rightarrow& e^{(a-\frac{1}{d})\frac{x'^2}{2g_s c} - \frac{rx'}{d} - \frac{g_s c r^2}{2d}} \,.
\ea
\paragraph{The $T$ transformation:} The kernel that follows for the transformation $T^k$ is
\ba
K_{T^k}(x',x) &=& e^{\frac{1}{2g_s k}  (x^2 - 2 xx' +  x'^2)}\,. \nn
\ea 
The result of the integration term by term is
\ba \label{TK}
\int dx \, K_{T^k}(x',x) e^{-rx} &\sim& q^{-k \frac{r^2}{2}} e^{-rx'} \,. 
\ea
Upon shifting $x$ by appropriate multiples of $g_s$ and $\pi i$, we thus find that the integral kernel $K_{T^k}$ implements framing shifts as expected,
\ba
Z^{(k)}(x') &=& \int dx \, K_{T^k}(x',x) Z(x - \frac{k}{2}g_s +k \pi i) \,.
\ea

\paragraph{The $S$ transformation:} The integration kernel $K_S$ is given by
\ba
K_S(x', x) &=& e^{\frac{1}{g_s} x x'} \,.\nn
\ea
This results in a Fourier transform. We can no longer apply the integral transform term by term to the partition function in series form. We can however evaluate the transformation when the partition function is presented as an infinite product, i.e. for the particular choice of framing that yields $Z^{1,2,3,4}$ in the form given in Eq. (\ref{productrep13}). Here, we will choose a contour in the complex $x$ plane so that we pick up all residues at $x=-t+g_s (i+\frac{1}{2})$, $i\ge0$ and $x=g_s (i+\frac{1}{2})$, $i\ge0$, respectively. We then obtain
\ba
\oint Z^{1,3}_t(x) e^{ x' x/g_s} &=& 
\oint \prod_{i=0}^\infty \frac{1- q^{i+ \frac{1}{2}} e^{-x}}{1-Q q^{i+\frac{1}{2}} e^{-x}} e^{x'x/g_s} \nn\\
&\rightarrow& (1-Q^{-1}) \prod_{i=1}^\infty \frac{1-Q^{-1} q^i}{1-q^{i}} \sum_{n=0}^{\infty} \prod_{i=1}^n \frac{1-Qq^i}{1-q^i} e^{(n+\frac{1}{2})x'}Q^{x'/g_s-n} \nn \\
&=&  \prod_{i=0}^\infty \frac{1-Q^{-1} q^i}{1-q^{i+1}} e^{\frac{g_s-2t}{2g_s} x'}  Z^{1,3*}_{t-g_s} (-x' -t +\frac{1}{2}g_s)
\ea
and
\ba
\oint Z^{2,4}_t(x) e^{x' x/g_s} &=& 
\oint \prod_{i=0}^\infty \frac{1- Qq^{i+ \frac{1}{2}} e^{-x}}{1-q^{i+\frac{1}{2}} e^{-x}} e^{ x' x/g_s} \nn\\
&\rightarrow& (1-Q) \prod_{i=1}^{\infty} \frac{1-Qq^i}{1-q^i} \sum_{n=0}^{\infty} \prod_{i=1}^{n} \frac{1-Qq^{-i}}{1-q^{-i}} e^{(n+\frac{1}{2})x'} \nn\\
&=&  \prod_{i=0}^{\infty} \frac{1-Qq^i}{1-q^{i+1}}  e^{\frac{1}{2}x'} Z^{2,4*}_{t+g_s}(-x'-\frac{1}{2}g_s) \,.
\ea
Note that $g_s$ shifts of closed string K\"ahler parameters have already made several appearances in studies of the open topological A-model, starting with \cite{DF}. Ignoring pre-factors and shifts, we interpret these equations schematically as
\ba
Z^{1,3}(x) &\xrightarrow{S}& Z^{1*,3*}(-x-t) \,, \nn\\
Z^{2,4}(x) &\xrightarrow{S}& Z^{2*,4*}(-x) \,.
\ea

The integral kernel for the transformation $S^2=-1$ cannot be read off of Eq. (\ref{intkernel}), as $c=0$. It is not hard however to convince oneself that the correct kernel here is simply
\ba
K_{S^2}(x', x) &=& \delta(x + x') \,.
\ea

We now check whether these results are consistent with the canonical coordinates determined at each puncture. The $SL(2,\IZ)$ matrices $A_{ij}$ mapping $(p_j,x_j)$ into $(p_i,x_i)$ can be read off from table \ref{concor}. The transformation matrix from $P_2$ to $P_1$ is given by
\ba
A_{12} &=&\left(
 \begin{matrix} 
      0 & -1 \\
      1 & 0 \\
   \end{matrix}
\right) = -S\,, \nn
\ea
and the transformation between $P_1$ and $P_3$, $P_2$ and $P_4$ respectively is affine linear, i.e. $A_{ij} (p_j, x_j)^T + a_{ij} = (p_i, x_i)^T$, with
\ba \nn
A_{13} = A_{24} =\left(
 \begin{matrix} 
      -1 & 0 \\
      0 & -1 \\
   \end{matrix}
\right) =  S^2  \,,&& 
a_{13} = \left(
\begin{matrix}
t \\
-t \\
\end{matrix} \right) \,, \hspace{0.3cm}
a_{24} = \left(
\begin{matrix}
-t \\
-t \\
\end{matrix} \right)  \,.
\ea
\begin{figure}[h]
\psfrag{A12}{${\scriptstyle{-S}}$}
\psfrag{A13}{${\scriptstyle{S^2 \circ \,+ \,(t,-t)^{T}}}$}
\begin{center}
\epsfig{file=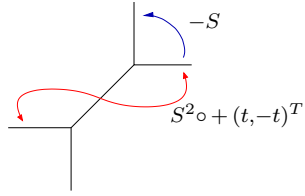, height=1in}
\end{center}
\caption{\small Canonical transformations relating different brane placements on the conifold.
\label{fig:cantransf}}
\end{figure}

\paragraph{Transformation from $P_1$ to $P_3$ and from $P_2$ to $P_4$:} This is given by $S^2=-1$ and shifts by $t$. Our transformation rules hence imply that the partition functions at the points of the two pairs should be related, respectively, by replacing the argument $x \mapsto -x -t$.
From our calculations above, the partition functions within each pair are equal. To check whether these two results are compatible, we must therefore relate the partition function at $x$ to the one at $-x$. To this end, note that at the points $P_2$ and $P_4$, the partition functions can be expressed as,
\ba
Z^{2,4} (x) &=& \frac{Z(x)}{Z(x+t)} \,, 
\ea
with 
\ba
Z(x) =\exp[ - \sum_{n=1}^{\infty} \frac{e^{-nx}}{n[n]}] \,.
\ea
$Z(x)$ is in fact the open topological string partition function for the vertex geometry with one brane insertion \cite{ADKMV}. Upon continuation away from real $g_s$, $Z(x)$ satisfies the following identity
\ba
Z^*(-x) &=& Z(x) \frac{q^{1/24}}{\eta(\tau)} \sum_{n=-\infty}^\infty (-1)^n q^{n^2/2} e^{-n x}  \,,  
\ea
with $q=e^{2\pi i \tau}$ and $x=-2\pi i\zeta$  (which is essentially the Jacobi triple product identity). We will take this as an indication for the relation
\ba 
Z^*(-x) \sim Z(x) \label{approx}
\ea
up to a normalization factor. Upon performing a modular transformation (again assuming continuation away from real $g_s$), this factor has an expansion in $e^{-1/g_s}$, suggesting that we interpret it as an indication for non-perturbative terms we are missing in the partition function. We will comment on this further in the next subsection.
Assuming (\ref{approx}), we obtain
\ba
Z^{2,4} (-x-t) &=& \frac{Z(-x-t)}{Z(-x)} \nn \\
&\sim& \frac{Z(x)}{Z(x+t)}=Z^{2,4}(x) \,.
\ea
Likewise for $Z^{1,3}$, 
\ba
Z^{1,3} (-x-t)  \sim  Z^{1,3} (x)  \,.\nn
\ea

\paragraph{Transformation from $P_2$ to $P_1$ and $P_4$ to $P_3$:} This is given by $- S$. It acts as
\ba
Z^{2,4}(x) \xrightarrow{S} Z^{2*,4*}(-x) = Z^{1,3}(-x) \xrightarrow{S^2} Z^{1,3}(x) \,. \nn
\ea
With the same caveats as above, the transformation kernel relating $Z^{2,4}$ to $Z^{1,3}$ hence follows correctly from the canonical variables assigned to $P_{1,3}$ and $P_{2,4}$. The inverse transformation from $P_{1,3}$ to $P_{2,4}$ is given by $(-S)^{-1} = S$. Here,
\ba
Z^{1,3}(x) \xrightarrow{S} Z^{1*,3*}(-x-t) =  Z^{2,4}(-x-t) \sim Z^{2,4}(x) \,. \nn
\ea

\subsection{Non-perturbative terms: a preliminary study} \label{s:non-per prelim}
In this final subsection, we offer some comments on the normalization factors that arise throughout the previous subsection.

In our study so far, we have followed the literature in assuming\footnote{In this subsection, we use conventions such that $x \rightarrow x + 2\pi$ is the periodicity of the exponential; the substitution $x \mapsto ix$ reinstates the conventions used in the rest of the paper. We will also express our results in terms of the Chern-Simons coupling $k$, rather than $g_s$. The two are related via $g_s = -\frac{2 \pi i}{k}$.}
\ba
\frac{\langle e^{ix} | e^{\oint_r A} | Z \rangle}{\langle e^{ix}  | Z \rangle  } &=& e^{irx} \,.  \label{lit}
\ea
On general grounds, normalized expectation values of Wilson loops in Chern-Simons theory on solid tori are known to yield characters of the relevant affine Lie algebra, in our case $\mathfrak{u}_k$. What has been neglected in (\ref{lit}) is to mod out by large gauge transformations (see \cite{marino}, which argues for this approach). If we instead take these into account, Eq. (\ref{lit}) is modified to
\ba
Z_r(x) &=&  \exp[i r x  ]  \sum_{m \in \IZ}  \exp[ik m x] \,, \label{modified}
\ea
which is indeed a $\mathfrak{u}_k$ character \cite{Elitzur}.

How does this modification effect our considerations? We consider here the transformation of the partition function $Z^*(x)$ on $\IC^3$, which was introduced in the previous subsection. The canonical transformation which maps different brane insertions amongst themselves is $-TS$ in this case \cite{ADKMV}. It acts as
\ba
Z^*(x) \xrightarrow{-TS}  Z(-x)  \,.
\ea
Recall that by 
\ba
Z^*(x) Z^*(-x) \sim \frac{\vartheta(x,1/k)}{\eta(1/k)}  \,,
\ea
we argued for $Z(-x) \sim Z^*(x) $.

The wave function for the antibrane with the modification (\ref{modified}), which we denote by $\tilde{Z}^*$, is given by
\ba
\tilde{Z}^*(x) &=& \sum_{r=0}^{\infty} \frac{1}{[r]!} q^{\frac{1}{4}(r-1)r} e^{irx} \sum_{m \in \IZ} e^{ikmx} \,.
\ea

Evaluating the Gaussian integral associated to the canonical transformation $-TS$ for the modified wavefunction term by term yields
\ba
\int dx \,K_{-TS} (x',x) e^{ix (r + km)} &=& e^{\frac{x'^2}{2 g_s} - i(r+km)x' + \frac{g_s (ir + ikm)^2}{2}} \\
&=& e^{-\frac{k x'^2}{4 \pi i}} e^{-i(r+km) x'} q^{-\frac{r^2}{2}} e^{\pi i  m^2 k} \,,
\ea
hence formally,
\ba
\tilde{Z}^*(x) \mapsto_{-TS}& & Z(-x-\pi) \, e^{-\frac{kx^2}{4 \pi i}} \vartheta(-\frac{kx}{2\pi},k)  \nn \\
&=& \sqrt{i/k} Z(-x-\pi)   \vartheta(\frac{x}{2 \pi},  -1/k) \nn \\
&=& \sqrt{i/k} \;\eta(-1/k) q^{-1/24}  \; Z^*(x) \,. \label{modtrans}
\ea
We see that the modification (\ref{modified}) yields precisely the $\vartheta$ function needed to relate $Z(-x)$ to $Z^*(x)$. Two main points remain to be understood:
\begin{itemize}
\item{The coefficient in (\ref{modtrans}), as well as the fact that we obtain $Z^*(x)$, and not $\tilde{Z}^*(x)$, are possibly due to our considering a too naive measure in performing canonical transformations (see \cite{Elitzur} for a discussion of the measure factor in the non-abelian context).}
\item{Our considerations regarding theta functions are formal as real $k$ corresponds to real modular parameter $\tau$, and this is a degenerate limit in which theta functions are ill-defined. This is yet another reflection of the fact that a complex variant of the conventional Chern-Simons theory is needed to capture the open topological string.}
\end{itemize}

\section*{Acknowledgments}
This work was partially supported by European RTN
under the contract 005104 ``Forces\-Universe'', and by  {\it l'Agence Nationale de la Recherche} under
the grants
ANR-06-BLAN-3$\_$137168 and ANR-05-BLAN-0029-01.


\begin{thebibliography}{0}
\bibitem{me}
  A.~K.~Kashani-Poor,
  ``The Wave Function Behavior of the Open Topological String Partition
  Function on the Conifold,''
  JHEP {\bf 0704}, 004 (2007)
  [arXiv:hep-th/0606112].
  %%CITATION = JHEPA,0704,004;%%

\bibitem{AV}
  M.~Aganagic and C.~Vafa,
  ``Mirror symmetry, D-branes and counting holomorphic discs,''
  arXiv:hep-th/0012041.
  %%CITATION = HEP-TH 0012041;%%  
  
\bibitem{hori}
  K.~Hori,
  ``Linear models of supersymmetric D-branes,''
  arXiv:hep-th/0012179.
  %%CITATION = HEP-TH 0012179;%%

\bibitem{KatzKlemm}
  S.~H.~Katz, A.~Klemm and C.~Vafa,
  %``Geometric engineering of quantum field theories,''
  Nucl.\ Phys.\  B {\bf 497}, 173 (1997)
  [arXiv:hep-th/9609239].
  %%CITATION = NUPHA,B497,173;%%

\bibitem{AKV}
  M.~Aganagic, A.~Klemm and C.~Vafa,
  ``Disk instantons, mirror symmetry and the duality web,''
  Z.\ Naturforsch.\ A {\bf 57}, 1 (2002)
  [arXiv:hep-th/0105045].
  %%CITATION = HEP-TH 0105045;%%

\bibitem{cswitten}
  E.~Witten,
  ``Chern-Simons gauge theory as a string theory,''
  Prog.\ Math.\  {\bf 133}, 637 (1995)
  [arXiv:hep-th/9207094].
  %%CITATION = HEP-TH 9207094;%%
  
\bibitem{Elitzur}
  S.~Elitzur, G.~W.~Moore, A.~Schwimmer and N.~Seiberg,
  ``Remarks On The Canonical Quantization Of The Chern-Simons-Witten Theory,''
  Nucl.\ Phys.\  B {\bf 326}, 108 (1989).
  %%CITATION = NUPHA,B326,108;%%

\bibitem{HV}
  K.~Hori and C.~Vafa,
  ``Mirror symmetry,''
  arXiv:hep-th/0002222.
  %%CITATION = HEP-TH 0002222;%%

\bibitem{HIV}
  K.~Hori, A.~Iqbal and C.~Vafa,
  ``D-branes and mirror symmetry,''
  arXiv:hep-th/0005247.
  %%CITATION = HEP-TH 0005247;%%

\bibitem{ADKMV}
  M.~Aganagic, R.~Dijkgraaf, A.~Klemm, M.~Marino and C.~Vafa,
  ``Topological strings and integrable hierarchies,''
  Commun.\ Math.\ Phys.\  {\bf 261}, 451 (2006)
  [arXiv:hep-th/0312085].
  %%CITATION = HEP-TH 0312085;%%

\bibitem{ANV}
  M.~Aganagic, A.~Neitzke and C.~Vafa,
  ``BPS microstates and the open topological string wave function,''
  arXiv:hep-th/0504054.
  %%CITATION = HEP-TH 0504054;%%

\bibitem{AKMV}
  M.~Aganagic, A.~Klemm, M.~Marino and C.~Vafa,
  ``The topological vertex,''
  Commun.\ Math.\ Phys.\  {\bf 254}, 425 (2005)
  [arXiv:hep-th/0305132].
  %%CITATION = HEP-TH 0305132;%%

\bibitem{vafanti}
  C.~Vafa,
  ``Brane/anti-brane systems and U(N$|$M) supergroup,''
  arXiv:hep-th/0101218.
  %%CITATION = HEP-TH 0101218;%%

\bibitem{katz}
  S.~Katz and C.~C.~Liu,
  ``Enumerative Geometry of Stable Maps with Lagrangian Boundary Conditions and Multiple Covers of the Disc,''
  Adv.\ Theor.\ Math.\ Phys.\  {\bf 5}, 1 (2002)
  [arXiv:math.ag/0103074].
  %%CITATION = MATH-AG 0103074;%%

\bibitem{DF}
  D.~E.~Diaconescu, B.~Florea and A.~Grassi,
  ``Geometric transitions and open string instantons,''
  Adv.\ Theor.\ Math.\ Phys.\  {\bf 6}, 619 (2003)
  [arXiv:hep-th/0205234].
  %%CITATION = HEP-TH 0205234;%%

\bibitem{marino}
  M.~Aganagic, A.~Klemm, M.~Marino and C.~Vafa,
  ``Matrix model as a mirror of Chern-Simons theory,''
  JHEP {\bf 0402}, 010 (2004)
  [arXiv:hep-th/0211098].
  %%CITATION = JHEPA,0402,010;%%



\end{thebibliography}
\end{document}